\documentclass[12pt,a4paper,dvips]{article}
\usepackage{epsfig,wrapfig,times,mathptm} 
\renewcommand{\section}[1]{\vspace{6pt} \noindent\mbox{#1} \newline \noindent}
\renewcommand{\subsection}[1]{\vspace{6pt} \noindent\mbox{\underline{#1}} 
\newline \noindent}
\renewcommand{\subsubsection}[1]{\vspace{6pt} \noindent\mbox{\underline{#1}}
\noindent}

\newfont{\sansb}{cmssbx10}
\newfont{\sans}{cmss10}

\def\gtrsim{\mathrel{\hbox{\rlap{\hbox{\lower4pt\hbox{$\sim$}}}\hbox{$>$}}}}

\begin{document}
{\small OG 4.3.3 \vspace{-24pt}\\}     
{\center \LARGE	TeV OBSERVATIONS OF THE VARIABILITY AND SPECTRUM OF 
MARKARIAN 501 
\vspace{6pt}\\}
J.~Quinn$^{1,2}$, 
I.H. Bond$^{3}$, 
P.J. Boyle$^{2}$, 
J.H. Buckley$^{1}$,  
S.M. Bradbury$^{3}$, 
A.C. Breslin$^{2}$,
A.M. Burdett$^{3}$, 
J. Bussons Gordo$^{2}$, 
D.A. Carter-Lewis$^{4}$, 
M. Catanese$^{4}$, 
M.F. Cawley$^{5}$,  
D.J. Fegan$^{2}$, 
J.P. Finley$^{6}$, 
J.A. Gaidos$^{6}$, 
A. Hall$^{6}$, 
A.M. Hillas$^{3}$, 
F. Krennrich$^{4}$, 
R.C. Lamb$^{7}$,  
R,W. Lessard$^{6}$,
C. Masterson$^{2}$, 
J.E. McEnery$^{2}$, 
G. Mohanty$^{4}$, 
P. Moriarty$^{8}$,
A.J. Rodgers$^{3}$, 
H.J. Rose$^{3}$, 
F.W. Samuelson$^{4}$, 
G.H. Sembroski$^{6}$, 
R. Srinivasan$^{6}$,
T.C. Weekes$^{1}$ 
and J. Zweerink$^{4}$\vspace{6pt}\\
{\it $^1$Whipple Observatory, Harvard-Smithsonian CfA, Box 97, Amado, 
AZ 85645, U.S.A.\\
$^2$Dept. of Experimental Physics, University College, Belfield, Dublin 4, 
Ireland \\
$^3$Physics Dept., University of Leeds, Leeds, LS2 9JT, Yorkshire, U.K.\\
$^4$Dept. of Physics and Astronomy, Iowa State University, Ames, 
IA 50011-3160, U.S.A.\\
$^5$Physics Dept., St. Patricks College, Maynooth, Ireland\\
$^6$Dept. of Physics, Purdue University, West Lafayette, IN 47907, U.S.A.\\
$^7$Space Radiation Lab., Caltech, Pasadena, CA 91125\\
$^8$Dept. of Physical Sciences, Regional Technical College, Galway, Ireland\vspace{-12pt}\\}

{\center ABSTRACT\\}
Markarian 501 is only the second extragalactic source to be
detected with high statistical certainty at TeV energies; it is
similar in many ways to Markarian 421. The Whipple Observatory
gamma-ray telescope has been used to observe the AGN Markarian 501
in 1996 and 1997, the years subsequent to its initial detection. 
The apparent variability on the one-day time-scale observed in TeV 
gamma rays in 1995 is confirmed and compared with the variability 
in Markarian 421. Observations at X-ray and optical wavelengths from 
1997 are also presented.

\setlength{\parindent}{1cm}
\section{INTRODUCTION}
Markarian 501 was discovered as a $\gamma$-ray source by the Whipple
Collaboration in 1995 (Quinn, et al., 1996) and has since been
verified by the HEGRA Collaboration (Bradbury et al., 1997). At the
time of its discovery the average emission level was 0.08 times that
of the Crab Nebula.  During 1995 the $\gamma$-ray emission from
Markarian 501 was observed to be constant, with the exception of one
occasion where the emission level rose to more than 5 standard
deviations above the average rate. Within 2 days the rate had returned
to its average level. This day-scale variability is also seen in
Markarian 421 (Kerrick et al., 1995). Markarian 501 is also seen to
vary at X-ray energies on time-scales from months (Mufson et al.,
1984) to hours (Giommi et al., 1990), and at optical wavelengths
significant variations as short as hours have been observed (Kidger
\& de Diego, 1992). The results of continuing $\gamma$-ray
observations made by the Whipple Collaboration in 1996 and 1997 are
presented below. Also presented are X-ray data from the
All-Sky Monitor of the Rossi X-ray Timing Explorer (2-10 keV) and
optical data taken (U-band) with the 1.2m telescope of the Whipple
Observatory during 1997.

The $\gamma$-ray observations reported in this paper were made with
the 10m Atmospheric \v{C}erenkov Imaging Telescope of the Whipple
Collaboration, which operates at an energy threshold of 350 GeV. The
telescope records images of the \v{C}erenkov radiation emitted from
cosmic ray initiated air-showers with a high resolution camera
consisting of a hexagonal array of photomultiplier tubes. Subsequent
analysis of the images determine which were probable $\gamma$-ray
events (Reynolds et al., 1993). Using this technique it is possible
to reject over 99.7\% of the background while retaining over 50\% of
the $\gamma$-ray events. The camera has recently been upgraded from
109 to 151 pixels (increasing the field of view from 3.0$^{\circ}$ to
3.5$^{\circ}$) and the $\gamma$-ray selection criteria were
re-optimised to take advantage of this. Therefore, to meaningfully
compare the data taken over the last 3 years the $\gamma$-ray rates
are expressed as a fraction of the rate from the Crab Nebula, which is
a steady source of TeV $\gamma$-rays (Hillas et al., 1997). This
approach assumes that the spectra of the objects are similar, an
assumption which initial investigations suggest is true (see below).

\begin{figure}[!ht]
 \vspace*{-0.3in}
 \centerline{\epsfig{figure=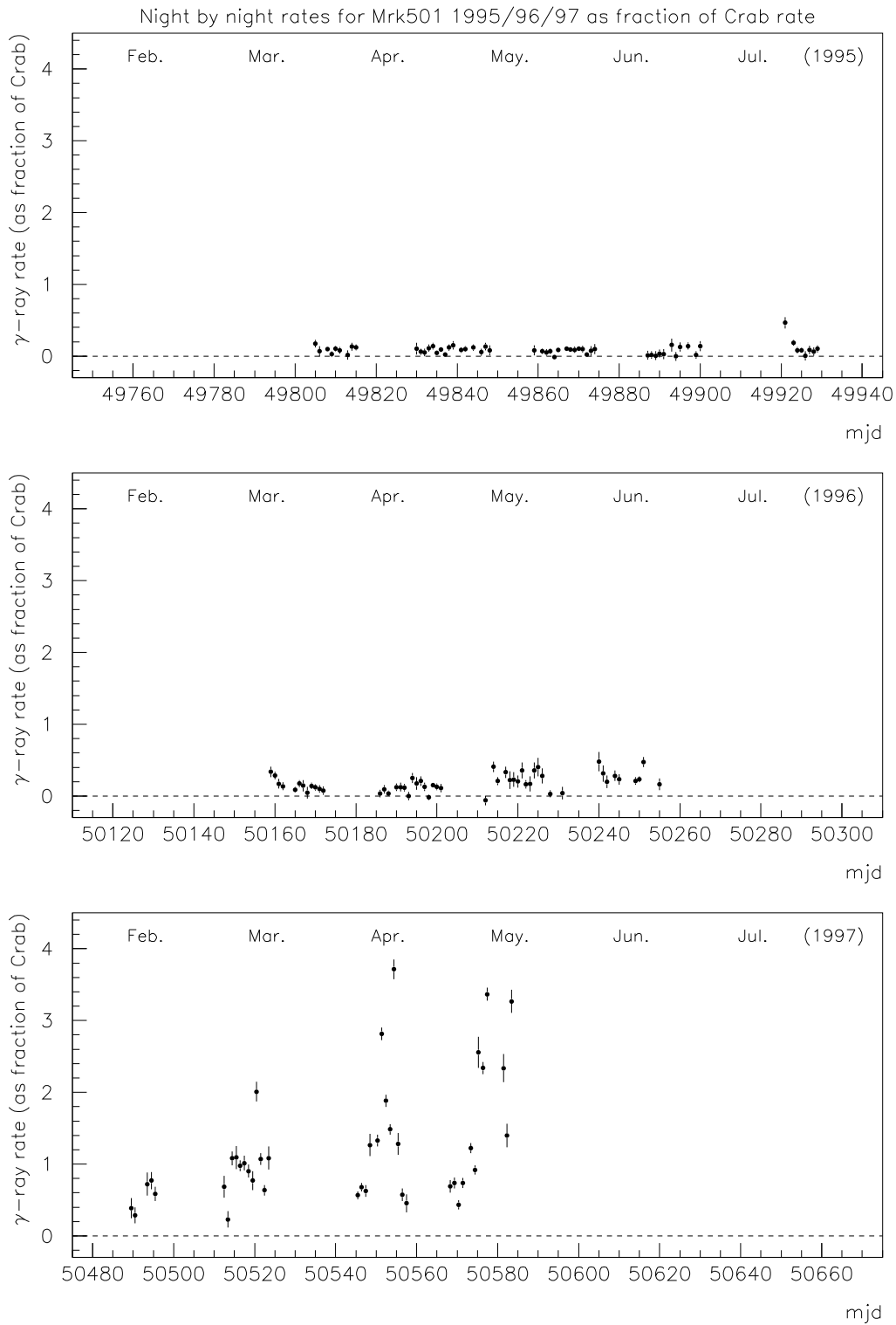,width=6.0in}}
 \vspace*{-0.3in}
 \caption{Daily $\gamma$-ray rates for Markarian 501 for last 3 years,
          expressed as fraction of average rate from Crab Nebula.}
 \label{fig:3yrs}
\end{figure}

\section{GAMMA-RAY FLUX VARIABILITY}
\begin{wrapfigure}[]{r}{4.6in}
 \vspace*{-0.3in}
\epsfig{figure=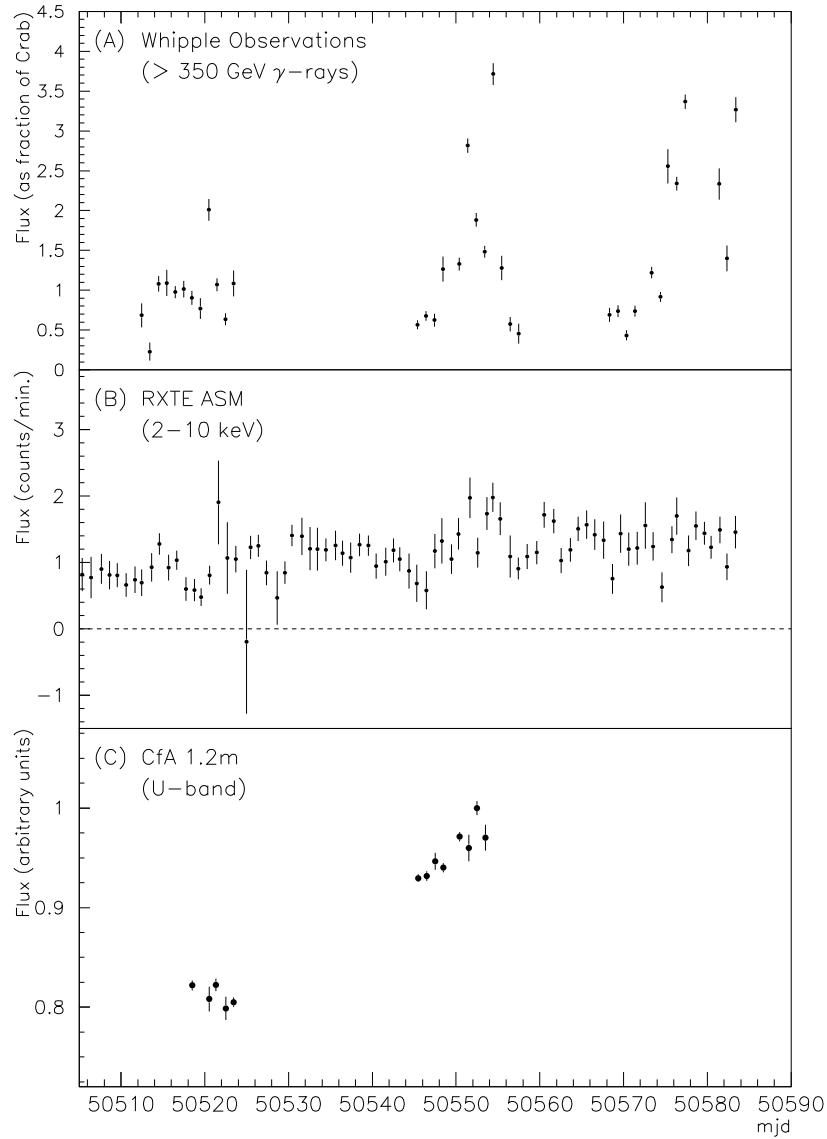,width=4.6in}
 \vspace*{-0.5in}
 \caption{Multi-wavelength observations of Markarian 501 in 1997:
          (a) Whipple ($>$ 300 GeV) $\gamma$-rays, (b) RXTE All Sky Monitor
	  and (c) 1.2m CfA (U-band).}
 \vspace*{-0.2in}
 \label{fig:multi}
\end{wrapfigure}

Figure \ref{fig:3yrs} shows the nightly flux levels for Markarian 501
for the 3 years of observations, up to and including May 1997. A subset of
the 1997 data is also shown in figure \ref{fig:multi}(a).  
As already stated, Markarian 501 exhibited variability on times-scales of
$\sim$1 day during 1995. The $\chi^{2}$ probability for the emission
being constant during those observations is 0.013 (0.38 if the night
of the flare is excluded). Observations in 1996 showed a doubling in
the average emission level, increasing from 0.08 times to 0.16 times
that of the average Crab level. There was no obvious flaring activity,
although the night-to-night rate did appear to be less stable. The
$\chi^{2}$ probability for constant emission for this year is
10$^{-12}$, indicating that there is variation on the scale of
days. The low signal-to-noise levels in these data do not permit the
search for variability on shorter time-scales.

A remarkable change in the $\gamma$-ray emission was seen in 1997: the
rate increased dramatically from previous seasons, the average now
being almost 1.3 times that of the Crab. There is also significant and
frequent flaring activity. On several occasions (May 12-13, April
12-13, April 15-16 and May 14-15) there have been flares with doubling
times of less than one day.  The $\chi^{2}$ probability for
constant emission is $<$10$^{-38}$ for this data while for Crab Nebula
data taken in 1997 it is 0.95. A preliminary analysis did not reveal
evidence for variability on time-scales less than 1 day, but a more
detailed study is underway.

\section{MULTI-WAVELENGTH OBSERVATIONS IN 1997}
In figure \ref{fig:multi}, flux measurements at VHE $\gamma$-ray, X-ray
and optical are presented. The X-ray data is taken from the Rossi
X-ray Timing Explorer (RXTE) All Sky Monitor data-set. It appears from
this data that Mrk501 is in a particularly active state, with a flux
higher than in previous years. The significance of the apparent
variations around the time of the April (MJD 50540 to 50560)
$\gamma$-ray data has not yet been established, nor has a correlation
analysis been performed. The optical data was taken with the 1.2m
telescope located at the Whipple Observatory.  For a more detailed
review of the optical data see Buckley \& McEnery (1997). The flux
(U-band) shows a significant increase ($\gtrsim$ 10\%) from March to
April while it appears to continue to increase during April.

\section{ENERGY SPECTRUM}
A spectral analysis of the 1997 Markarian 501 data, using the method
of Mohanty (1997) is currently under way and details will be presented
at this conference. However, it is possible to get an indication of
how the energy spectrum compares to that of the Crab Nebula by
examining size (total signal in d.c. in an image) distributions of the
candidate $\gamma$-ray events. The method of Vacanti et al. (1991) was
used to calculate the size spectra and preliminary results indicate
that the spectrum of Markarian 501 may be slightly harder than that of
the Crab Nebula; the spectra are not, however, significantly different
when systematic effects are taken into consideration. Large zenith
angle observations indicate that the spectrum extends out to at least
7 TeV (Krennrich, 1997).

\section{DISCUSSION}
A remarkable change in the TeV $\gamma$-ray emission from Markarian
501 has been seen in 1997. The average emission level has increased by
a factor of more than 16 since its discovery as a $\gamma$-ray source
in 1995. Day-scale variability, as we have seen on one occasion in
1995, has been confirmed and the frequency of the flaring appears to
have increased. Day-scale variability is a property which Markarian
501 has in common with Markarian 421 (Buckley et al., 1996). However,
preliminary analysis of the data does not reveal any evidence for
hour-scale variability, as is seen in Markarian 421 (Gaidos et al.,
1996). The $\gamma$-ray emission from Markarian 501 rarely goes to
zero, in contrast to Markarian 421, whose flux has been described as
being composed of a series of rapid flares, with no underlying
baseline emission (Buckley et al., 1996). Detailed analyses to search
for short ($<$ one day) term variability, to determine an energy
spectrum and to look for possible correlations with other wavelength
data are currently underway.

\section{ACKNOWLEDGEMENTS}
This research is supported by grants from the U.S. Department of Energy,
by NASA, by PPARC in the U.K. and by FORBAIRT in Ireland. The X-ray
data in this work are quicklook results presented by the ASM/RXTE team.

\section{REFERENCES}
\setlength{\parindent}{-5mm}
\begin{list}{}{\topsep 0pt \partopsep 0pt \itemsep 0pt \leftmargin 5mm
\parsep 0pt \itemindent -5mm}
\vspace{-15pt}
\item Bradbury, S.M., et al., A\&A, 320, L5, (1997).
\item Buckley, J.H, and McEnery, J.E., in preparation, (1997).
\item Gaidos, J.A. et al., Nature, 383, 319, (1996).
\item Hillas, A.M. et al., in preparation, (1997).
\item Kerrick, A.D., et al. ApJ, 438, L59, (1995).
\item Kidger, M.A. \& de Diego, J.A., A\&AS, 93, 1, (1992).
\item Krennrich, F., et al., to be published in Towards a Major Atmospheric \v{C}erenkov Detector V, (1997). 
\item Mohanty, G. et al., in press, (1997).
\item Mufson, S.L. et al., ApJ, 285, 571, (1984). 
\item Quinn, J., et al., ApJ, 456, L83, (1996).
\item Reynolds, P.T., et al., ApJ, 404, 206, (1993).
\item Vacanti, G., et al., 377, 467, (1991).

\end{list}

\end{document}